\documentclass[preprint,12pt]{elsarticle}

\usepackage{hyperref}
\hypersetup{colorlinks=true, allcolors=black}
\usepackage{graphicx}
\usepackage{amsmath} 
\usepackage{booktabs}

\usepackage{algorithm} 
\usepackage{algpseudocode} 
\usepackage{caption}
\usepackage{setspace}

\usepackage[margin=2.5cm]{geometry}

\usepackage{amssymb,amsthm}
\theoremstyle{definition}
\renewcommand{\qedsymbol}{$\square$} 

\usepackage{algorithm}
\usepackage{algorithmicx}
\usepackage{algpseudocode}
\usepackage{caption}
\algnewcommand\Input{\item[\textbf{Input:}]}
\algnewcommand\Output{\item[\textbf{Output:}]}

\usepackage{bbm}
\usepackage{bm}

\newcommand{\bbE}{{\mathbb E}}

\newcommand{\bX}{{\bm X}}

\newcommand{\cN}{{\mathcal N}}

\newcommand{\cV}{{\mathcal V}}
\newcommand{\cE}{{\mathcal E}}
\newcommand{{\cF}}{{\mathcal F}}
\newcommand{\cL}{{\mathcal L}}

\usepackage{pdflscape}
\usepackage{float}
\usepackage{array}

\graphicspath{{figures/}}

\journal{Entropy}

\begin{document}

\begin{frontmatter}



\title{Reconstructing Sparse Multiplex Networks with Application to
Covert Networks}


\author[address1]{Jin-Zhu Y\"{u}}
\author[address2]{Mincheng Wu}
\author[address3]{Gisela Bichler}
\author[address4]{Felipe Aros-Vera}
\author[address5,address6]{Jianxi Gao\corref{label1}}
\ead{jianxi.gao@gmail.com}

\cortext[label1]{Corresponding author}
\address[address1]{ Department of Civil Engineering, University of Texas at Arlington, Arlington, TX, USA}
\address[address2]{College of Control Science and Engineering, Zhejiang University, Hangzhou, Zhejiang, China}
\address[address3]{Department of Criminal Justice, California State University, San Bernardino, CA, USA}
\address[address4]{Department of Industrial and Systems Engineering, Ohio University, Athens, OH, USA}
\address[address5]{Department of Computer Science, Rensselaer Polytechnic Institute (RPI), Troy, NY, USA}
\address[address6]{Network Science and Technology Center, Rensselaer Polytechnic Institute (RPI), Troy, NY, USA}
\begin{abstract}
Network structure provides critical information for understanding the dynamic behavior of networks. However, the complete structure of real-world networks is often unavailable, thus it is crucially important to develop approaches to infer a more complete structure of networks. In this paper, we integrate the configuration model for generating random networks into an Expectation-Maximization-Aggregation (EMA) framework to reconstruct the complete structure of multiplex networks. 
We validate the proposed EMA framework against the random model on several real-world multiplex networks, including both covert and overt ones. It is found that the EMA framework generally achieves the best predictive accuracy compared to the EM framework and the random model. As the number of layers increases, the performance improvement of EMA over EM decreases. The inferred multiplex networks can be leveraged to inform the decision-making on monitoring covert networks as well as allocating limited resources for collecting additional information to improve reconstruction accuracy. For law enforcement agencies, the inferred complete network structure can be used to develop more effective strategies for covert network interdiction.
\end{abstract}


\end{frontmatter}



\section{Introduction}
\label{sec:intro}
\begin{figure}[!h]
    \centering
    \includegraphics[width=0.6\textwidth]{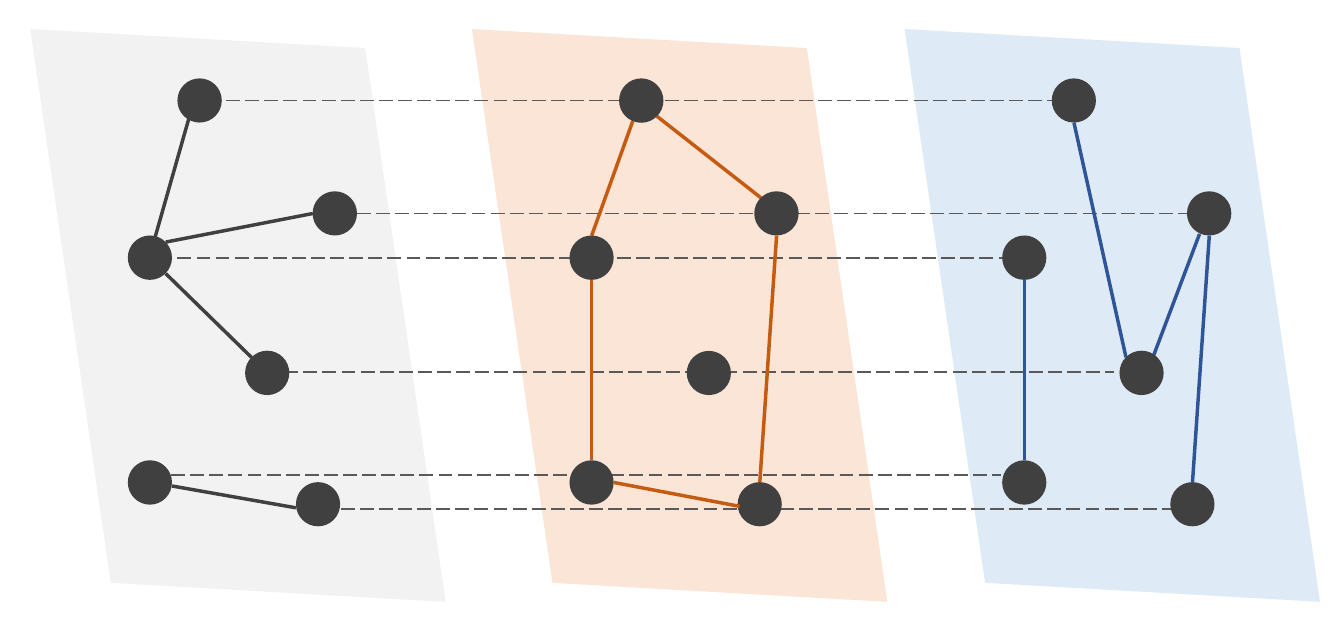}
    \caption{A schematic of a multiplex network with three layers and {six} nodes. In a multiplex network, the nodes are shared across layers, and links in each layer represent one type of relation between nodes. An interlayer link exists between any node and its counterparts in other layers.}
    \label{fig:multiplex_eg}
\end{figure}

Multiplex networks, in which each layer consists of the same set of nodes but different sets of links (Fig. \ref{fig:multiplex_eg}), are a powerful tool for describing and analyzing the connectivity from interactions of different types among the same set of entities in complex systems\cite{gao2022introduction}. Examples include social networks \cite{zhu2019information}, multi-modal transportation networks \cite{sole2016congestion}, multi-tier supply networks \cite{kim2019exploring}, etc. 
Due to the interplay between network structure and dynamics, the knowledge of complete multiplex network structure is essential to a deep understanding of the dynamic behavior of multiplex networks and predicting future interactions between nodes in the networks \cite{aleta2020link}. Nonetheless, in many practical applications, particularly covert networks, it is extremely difficult to obtain sufficient data for constructing the complete structure of all layers of a multiplex network due to limited observing resources and privacy concerns \cite{tran2020deepnc}.

Reconstructing the {complete} topology {(i.e., all nodes and links)} from incomplete {toplogies (partially observed nodes and links)} is an exceedingly challenging task primarily because (i) the number of missing nodes and links will be very large, (ii) sparse networks (e.g., illicit supply networks) usually have much lower link density than typical (overt) social networks, thereby it is often inappropriate to use many prediction methods developed for denser networks with a balanced ratio of positive and negative labels, and (iii) observations are limited to task-specific activity, excluding historic relations that are instrumental to network resilience \cite{gao2016universal,liu2020network}.


On the application side, this study focuses on illicit networks, such as drug and human trafficking networks. The reason is that illicit networks compromise national security and prosperity as well as public health and safety \cite{xu2008topology}. Despite considerable advances in understanding and disrupting the operations of covert networks, a deep understanding of the structure and dynamics of covert networks remains lacking. Although the structure is crucial for understanding, monitoring, and controlling covert networks, this topological information is always incomplete \cite{hosseinkhani2012discovering, pourhabibi2021darknetexplorer}. 
There are multiple reasons: (i) Investigations typically focus on current operations with the aim of documenting the involvement of specific individuals in specific criminal activities. (ii) Potential criminal actors may be identified and are then often omitted from the network due to essential legal principles and civil rights protections. (iii) Criminal actors take measures to obfuscate their involvement, hiding, or disguising activities and relations because they are wary of attack by agents of the criminal justice system, as well as competitors or rivals. 
By modeling illicit networks as a multiplex and reconstructing the structure of multiplex covert networks, this study can be leveraged to help monitor, control, and interdict the illicit activities on covert networks.

The rest of this paper is structured as follows. Section 2 presents the literature review. Details on the EMA framework are presented in Section 3. Details on the real-world multiplex networks used to validate the proposed EMA framework are given in Section 4. The numerical experiments and the resultant results are reported in Section 5, followed by the discussion on the implication of this study on disrupting illicit networks in Section 6 and the concluding remarks in Section 7.

\section{Literature Review}

Network reconstruction methods can be divided into three categories: similarity-based approach, probabilistic (statistical) approach, and algorithmic (machine learning) approach. 

Structural similarity-based methods assume that the nodes tend to connect to other nodes with a higher level of similarity computed by a certain distance
function \cite{martinez2016survey}. These approaches give similarity scores for missing or unobserved links between any pair of two nodes, then links with high similarity scores will be predicted to exist. For example, in \cite{abdolhosseini2020link}, the eigenvectors of the layer adjacency matrix are used to measure the layer topological similarity and then the topological similarity (element of the layer similarity matrix) of the unconnected nodes is used to predict if a link exists between them. \citet{berlusconi2016link} and \citet{calderoni2020robust} apply multiple similarity metrics, such as Common Neighbor and Resource Allocation, to identify missing links in a criminal network. They demonstrated that these similarity metrics can identify possible missing links in criminal networks with noise or incomplete information. However, the performance of similarity-based methods depends on the structure of specific networks \cite{lu2011link}, therefore prior knowledge about the networks under study is usually required to improve the predictive performance \cite{abdolhosseini2020link}. This approach is infeasible for the network reconstruction problem in this study because the nodes and links to be inferred {are completely missing without} any connection with observed nodes, therefore all scores of {structural} similarity between unobserved nodes and observed nodes are zeros.

Statistical approaches for network structure inference assume a network has a known structure, and then a parametric or nonparametric model is built to fit the structure, such as the stochastic block model or its variants, as well as Dirichlet network distribution model \cite{herlau2014infinite,williamson2016nonparametric, stanley2019stochastic,wu2021quantifying}. Then the model parameters are estimated using statistical methods, such as maximum likelihood estimation. The model
parameters are then used to compute the formation probability of each unobserved link, such as the stochastic block model and the hierarchical structure model. In particular, \citet{kim2011network} develop a scalable Kronecker Expectation-Maximization (KronEM) approach to reconstruct a single-layer network with a known number of missing nodes by alternating between inferring the unobserved part of the network (E-step) and estimating the optimal model parameters (M-step) until convergence. The network structure is assumed to be described by the Kronecker graph model, and this approach does not require attributes about nodes or edges. \citet{wu2022discrimination} use a Bayesian approach to reconstruct two-layer multiplex networks from the available aggregate topology of all layers (the aggregation of all the layers into a monoplex layer) and partial observations of nodes in each layer, and the inference is conducted by leveraging the Expectation-Maximization algorithm.
However, statistical approaches would be computationally challenging for large networks, especially when a large number of samples are required for inference \cite{kim2011network}.

The previous two approaches are based on computing a score for each unobserved link by defining a similarity or a probability function. However, the inference of network structure can also benefit from other algorithmic (machine learning) approaches, including supervised learning and optimization techniques. In particular, \citet{zhang2018link} develop graph neural networks (GNN) models to predict missing links. However, the method is mainly developed for single-layer networks.
\citet{gao2022inductive} develop a network embedding method called interactive learning across relations that exploits existing multiple types of relational data to predict inductive links (new  or unobserved relation) between existing nodes instead of inter- or intra-layer links in a known layer.

On the application side, a growing number of researchers are resorting to multilayer network models to study illicit supply chains~\cite{pourhabibi2021darknetexplorer,baycik2018interdicting,kosmas2020interdicting,shen2021interdicting,ficara2021multilayer} since these models allow actors to be placed in different layers representing different characteristics of covert networks. Going beyond mapping only the co-offending activity, multilayer models help to reveal more about the social processes that connect people, which can be used to infer missing information~\cite{traganitis2017topology, chen2017cross}.
However, to the best of our knowledge, a few studies have attempted to infer their multiplayer structure (topology), even though the complete structure of multiplex illicit networks is typically missing, as discussed earlier.

In this study, we develop an EMA framework built on the EM algorithm. The EMA framework is advantageous for reconstructing covert networks because: 

\begin{enumerate}
    \item The sparsity of links in covert networks, i.e., the negatives (no links between nodes) significantly outnumber the positives, leads to challenges in inference and thus significantly affects the predictive performance of classical learning-based method \cite{wang2021distance}.
    \item It can infer the links that connect unobserved nodes of the network. This is hardly possible for methods built on structural similarities, such as Adamic-Adar index, Jaccard index, and Resource Allocation index {\cite{liben2003link}}, because the links to be inferred usually connect unobserved nodes that are isolated from the observed part of the network. 
    \item It embeds observed criminal activity within the latent social structure required to sustain operations within a dynamic operational context \cite{agreste2016network, smith2016trust,malm2013using, kleemans1999social, duijn2014relative}. This informs disruption efforts by improving estimates of network resiliency \cite{krebs2002terror, malm2010comparing}. 
\end{enumerate}

The EMA framework is an extension to the EM framework proposed in a recent work \cite{wu2022discrimination}, but the EMA framework is more generic because it can be applied to multiplex networks with more than two layers and unknown aggregate topology, which is more common for real-world incomplete multiplex networks.



\section{Methodology} \label{sec:method}
\noindent
\subsection{Problem Description}
Mathematically, a multiplex network can be described by a set {$\bm Z = \{\bm Z^{\ell} {|}\, \ell=1,2,\cdots,m \}$}, where $m$ is the number of layers. $\bm Z^{\ell}$ consists of the {complete} set of nodes $\cN$ and the {complete} set of edges $\cE^{\ell}$ of relation type ${\ell}$. 
For incomplete multiplex networks, only some parts ({partial topologies}) of each layer are observed, i.e., a fraction $c$ of nodes and the associated links of the complete network. {If we denote the {partially observed topologies} as $\bm X = \{ \bm X^{\ell} {|}\, \ell=1,2,\cdots,m\}$, then we have $\bm X^{\ell} \subseteq \bm Z^{\ell}$.} 
{In this study, we use the same notation for topology and the adjacency matrix that encodes the connectivity of the topology. We consider undirected and unweighted multiplex networks, therefore the observed and complete adjacency matrices for each layer of a multiplex network are symmetric and binary (i.e., $\bm Z_{ij}^{\ell}, \bm X_{ij}^{\ell} \in \{0,1\},\, \forall i, j \in \cN$).} Formally, the problem of this study can be given as follows.\\

\noindent
\textbf{Problem.} \textit{Consider an {undirected and unweighted} multiplex network wherein each layer $\ell$ have the same set of nodes $\cN$ but different set of links {$\cE^{\ell}$}. The multiplex network reconstruction problem is to infer the complete topology of all layers {$\bm Z$} given {partial topologies} of all layers {$\bm X$} }.\\


\noindent
Note that the {classical} link prediction problem {\cite{liben2003link}}, in which only links are missing { and typically not all links connected a node are missing}, is a special case of the network reconstruction problem, in which both nodes and links can be missing, {i.e., all links connected to a missing node are missing as well}. As such, the network reconstruction problem is naturally more difficult than the link prediction problem.




\begin{table}[!tb]
\begin{tabular}{lp{11.9cm}l}
\midrule
\multicolumn{2}{c}{\textbf{Notations}} \\
\multicolumn{2}{l}{\textbf{Sets \& Indices}} \\
$i,\;j$ & Indices of nodes\\
$\ell$ & Index of layers\\
$\cL$ & Set of layers\\
$\cN$, $\cE$ & Sets of nodes and links\\
$\cV_{obs}$ & Set of observed nodes\\
{$\bm X$} & Set of the observed {topologies} (by layer) of a multiplex network\\
{$\bm Z$} & Set of complete {topologies} (by layer) of a multiplex network\\
\multicolumn{2}{l}{\textbf{Variables \& Parameters}} \\
$c$  &  Fraction of observed nodes in a layer\\
$\bm d$ &  Degree sequences\\
{$\langle d \rangle$} & Average degree\\
$m$ &  Number of layers\\
$U,\,V$ &  Generic random variables\\
$\bm X^{\ell}$ & Observed adjacency matrix (subnetwork) for layer $\ell$\\
$\bm Z^{\ell}$ & Adjacency matrix (complete topology) for layer $\ell$\\
$\bm A$ & Adjacency matrix for the aggregate topology of a multiplex network\\
$|CC|$, $|GCC|$ & Size of a connected component and size of the greatest connected component\\
{$\langle |CC| \rangle$} & Average size of connected components\\
$\epsilon$, $\epsilon_T$ & Convergence error and error tolerance\\
{$p_{ij}$} & Probability of a link between node $i$ and node $j$.\\
$q(\bm Z)$ & Posterior distribution for the complete multiplex network\\
$\rho$ & Link density\\
$\bm{\Theta}$ & Model parameters\\
\midrule
\end{tabular}
\end{table}


















\subsection{Expectation-Maximization-Aggregation Framework}
This section provides details on the proposed EMA framework for multiplex network reconstruction, including the classical EM algorithm, the network generation model, the integration of aggregate topology into the EM framework, as well as the corresponding algorithms. 

\subsubsection{Expectation-Maximization}
The EM algorithm \cite{dempster1977maximum} is an iterative approach for estimation problems wherein the value of model parameters and latent variables depend on each other. For the network reconstruction problem, where the task is to find the most probable topology of the missing subnetwork, it is required to first estimate the parameters of the network model, which establish a connection between the observed subnetworks $\bm X$ and the {complete multiplex network} $\bm Z$. This framework can be formulated as follows. 

We here consider a model parameterized by $\bm \Theta$. Given the observed subnetworks $\bm X$ of the complete multiplex network $\bm Z$, the task of classical maximum likelihood estimation is to estimate $\bm \Theta$. However, since the complete observations $\bm Z$ are not available, we need to marginalize it out. Let $q(\bm Z)$ be the probability density distribution of $\bm Z$, i.e., ${\sum_{\bm Z}} q(\bm Z) =1$ and $q(\bm Z) \ge 0$, 
then we get the likelihood of observed topology given the model 

\begin{subequations}
\begin{align}
    \ln p\left( \bm X \mid \bm \Theta \right) 
    &= \ln {\sum_{\bm Z}} p\left( \bm X, \bm Z \mid \bm \Theta \right) \label{eq:orig_likeli}\\
    &= \ln {\sum_{\bm Z}} q(\bm Z) \cdot \frac{p\left( \bm X, \bm Z \mid \bm \Theta \right)}{q(\bm Z)}\\
    & \ge  {\sum_{\bm Z}} q(\bm Z) \ln \left(\frac{p\left( \bm X, \bm Z \mid \bm \Theta \right)}{q(\bm Z)} \right).\label{eq:elbo}
\end{align}
\end{subequations}



\noindent
The last inequality follows from Jensen's equality. Specifically, $\ln \bbE[x] \ge \bbE[\ln x]$ since $\ln x$ is a concave function. The equality is attained when $\frac{p\left( \bm X, \bm Z \mid \bm \Theta \right)}{q(\bm Z)}$ is a constant, i.e., $q(\bm Z) \propto p\left( \bm X, \bm Z \mid \bm \Theta \right)$. Typically, $q(\bm Z)$ is set to the posterior distribution of $\bm Z$, thus we get

\begin{equation} \label{eq:post_Z}
  q(\bm Z) = \frac{p\left( \bm X, \bm Z \mid \bm \Theta \right)}{{\sum_{\bm Z}} p\left( \bm X, \bm Z \mid \bm \Theta \right) }= p(\bm Z\mid \bm X, \bm \Theta)  .
\end{equation}

The lower bound given by Eq.~\eqref{eq:elbo}, referred to as the evidence lower bound (ELBO), is used because it is difficult to directly optimize Eq.~\eqref{eq:orig_likeli} due to the missing information \cite{blei2017variational}. The EM algorithm then proceeds by alternating between the following two steps until convergence \cite{dempster1977maximum}:

\begin{enumerate}
    \item E-step: updating $q(\bm Z)$ by setting it to the posterior of $\bm Z$ given the current $\bm \Theta$, which is used to construct the local ELBO on the log-likelihood.
    \item M-step: updating $\bm \Theta$ by maximizing the ELBO, i.e., solving for $\bm \Theta$ in 
    
    \begin{equation}
       \frac{\partial}{\partial \bm \Theta} {\sum_{\bm Z}} q(\bm Z) \ln p\left(\bm X, \bm Z \mid \bm \Theta\right) =0.
    \end{equation}

\end{enumerate}

\subsubsection{Network Generation Model}
In this study, the parameters in the model for generating the topology of each layer include the degree sequences of layers $\bm d$ and the adjacency of the aggregate topology from all layers of the multiplex network $ \bm A$. The degree sequence of a layer is used to generate the topology of that layer using the configuration model \cite{bender1978asymptotic,newman2018networks}. According to this model, for a network with a given degree sequence, the probability of a link between two nodes with degree $\bm d_i$ and $\bm d_j$ is $\frac{\bm d_i \bm d_j}{2|\cE|-1}$ where $|\cE|$ is the number of edges. Therefore, after estimating the degree sequences of all layers, we can probabilistically infer the complete topology of each layer. The adjacency matrix of the aggregate topology is included in the model parameters because we add an aggregation step to the EM framework to obtain the aggregated topology of all layers. In this way, information on the interlayer dependence of the multiplex network is leveraged to improve the reconstruction accuracy. Without this step, the topologies of layers are independent of each other and are hence reconstructed separately.

\subsubsection{Full Algorithm for the EMA Framework}

This section presents details about how to reconstruct multiplex networks using the EMA framework, including the E-step, M-step, and A-step. 

In the E-step, the task is to obtain the posterior of $\bm Z$ given by Eq. \eqref{eq:post_Z}. 
According to the configuration model, 
{the probability of a link between nodes $i$ and $j$ in layer ${\ell}$ before it is not observed yet} can be given by

{
\begin{equation} \label{eq:Z_orig}
    p_{ij}^{\ell} = 
    \frac{\bm d_i^{\ell} \bm d_j^{\ell}}{2 {|\cE^{\ell}|} - 1}
\end{equation}
}

\noindent
Considering that $\frac{\bm d_i^{\ell} \bm d_j^{\ell}}{2 |\cE^{\ell}| - 1}$ can be greater than 1 and certain entries are observed, Eq. \eqref{eq:Z_orig} is changed to 

{
\begin{equation} \label{eq:Z_final}
  p_{ij}^{\ell} =
  \begin{cases}
    \bX_{ij}^{\ell}, & \text{if } i \in \cV_{\text{obs}} \, \text{and} \, j \in \cV_{\text{obs}} \\
    \min (1, \frac{\bm d_i^{\ell} \bm d_j^{\ell}}{2 {|\cE^{\ell}|} - 1}), &  \text{otherwise}
  \end{cases}
\end{equation} 
}


\noindent
{Then $q \left(\bm Z_{ij}^{\ell} \right)$ is expressed as}

{ 
\begin{equation}
 q(\bm Z_{ij}^{\ell}) = (p_{ij}^{\ell})^{\bm Z_{ij}^{\ell}} \left(1-p_{ij}^{\ell}\right)^{1-\bm Z_{i j}^{\ell}},\, \bm Z_{ij}^{\ell}\in\{0,1\},  
 \label{eq:q_Z} 
\end{equation} 
}

Next in the M-step, {given the current $q \left(\bm Z_{ij}^{\ell} \right)$}, we solve for $\bm d^{\ell}_i$ for any $i \in \cN,\, \ell \in \cL$ as follows. 
In the ELBO, the term involving $\bm d^{\ell}_i$ is

\begin{align}\label{eq:di_in_elbo}
{ \sum_{\bm Z_{ij}^{\ell}} } q(\bm Z_{i j}^{\ell})  \ln  \left( \frac{\left(\frac{\bm d^{\ell}_i   \bm d_j^{\ell}}{2|\cE^{\ell}|-1}\right)^{\bm Z_{i j}^{\ell}} \left(1-\frac{\bm d^{\ell}_i   \bm d_j^{\ell}}{2|\cE^{\ell}|-1}\right)^{1-\bm Z_{i j}^{\ell}}}{q(\bm Z_{i j}^{\ell})} \right) 
\end{align}

\noindent 
To simplify the term involving $\bm d_i^{\ell}$ in Eq. \eqref{eq:di_in_elbo}, we let $a_j=\frac{\bm d_j^{\ell}}{2|\cE^{\ell}|-1}$. Then differentiate with respect to an arbitrary $\bm d_i^{\ell}$ and we obtain

\begin{subequations}
\begin{align}
  &{ \sum_{\bm Z_{ij}^{\ell}} }\frac{\partial}{\partial \bm d^{\ell}_i} \left( q(\bm Z_{i j}^{\ell}) \cdot \ln \left[ ( a_j \bm d^{\ell}_i ) ^{\bm Z_{i j}^{\ell}} (1- a_j \bm d^{\ell}_i)^{1-\bm Z_{i j}^{\ell}} \right] \right) \\
  =&  { \sum_{\bm Z_{ij}^{\ell}} } q(\bm Z_{i j}^{\ell}) \cdot \frac{\bm Z_{i j}^{\ell}-a_j \bm d^{\ell}_i}{\bm d^{\ell}_i (1-a_j \bm d^{\ell}_i)} \\
  =& \frac{\bbE\left[\bm Z_{i j}^{\ell}\right] - a_j \bm d^{\ell}_i}{{\bm d^{\ell}_i (1-a_j \bm d^{\ell}_i)}} 
    \label{eq:diff_d_i}
\end{align}
\end{subequations}

%
\noindent
Let Eq.\;\eqref{eq:diff_d_i} = 0 and we get 

\begin{equation}
    { \bm d^{\ell}_j \bm d^{\ell}_i = \left({2 |\cE^{\ell}| - 1}\right) {\bbE\left[\bm Z_{i j}^{\ell}\right]}, \; \forall i \in \cN.}
\end{equation}

\noindent
{To ensure that the solution to $\bm d^{\ell}_i$ does not involve $\bm d^{\ell}_j$ , we sum both sides of the above equation over $j$ and obtain}

\begin{equation}
    \bm d^{\ell}_i =\frac{{2 |\cE^{\ell}| - 1}}{{2|\cE^{\ell}|}-\bm d^{\ell}_i} \sum_{j \in \cN} \bbE\left[\bm Z_{i j}^{\ell}\right] \label{eq:inter_sol_di}, \; \forall i \in \cN.
\end{equation}

\noindent
Solving for $\bm d^{\ell}_i$, we have

\begin{equation}
     \bm d^{\ell}_i = |\cE^{\ell}| - \sqrt{|\cE^{\ell}|^2 - 2|\cE^{\ell}| \sum_{j \in \cN} \bbE\left[\bm Z_{i j}^{\ell}\right] + \sum_{j \in \cN} \bbE\left[\bm Z_{i j}^{\ell}\right] }.
\end{equation}

\noindent
{Note that Eq. \eqref{eq:inter_sol_di} has two solutions to $\bm d^{\ell}_i$, but the other solution is discarded because it is greater than the number of links in layer $\ell$.}
For $\bm d^{\ell}_i$ to have a real-valued solution, it is easy to show that the following condition should be satisfied

{
\begin{equation}
    \left(|\cE^{\ell}|-\sum_{j \in \cN} \bbE\left[\bm Z_{i j}^{\ell}\right]\right)^2 \ge  \left(\sum_{j \in \cN} \bbE\left[\bm Z_{i j}^{\ell}\right] \right)^2 - \sum_{j \in \cN} \bbE\left[\bm Z_{i j}^{\ell}\right], \; \forall i \in \cN,
\end{equation} 
}

\noindent
{Since $|\cE^{\ell}|-\sum_{j \in \cN} \bbE\left[\bm Z_{i j}^{\ell}\right] \ge 0$, we have
}

\begin{equation}
    \frac{|\cE^{\ell}|}{\sum_{j \in \cN} \bbE\left[\bm Z_{i j}^{\ell}\right] } \ge  \sqrt{\frac{\sum_{j \in \cN} \bbE\left[\bm Z_{i j}^{\ell}\right]-1}{\sum_{j \in \cN} \bbE\left[\bm Z_{i j}^{\ell}\right] }} + 1, \; \forall i \in \cN,
\end{equation} 

\noindent which indicates that a layer should not have giant hub nodes that are incident with over 50\% of the edges of a layer. This condition is typically satisfied for sparse multiplex networks. For layers with a sufficiently large number of edges but no giant hub node, $|\cE^{\ell}| \gg \bm d^{\ell}_i$, thus the solution to Eq.~\eqref{eq:inter_sol_di} can be approximately given by

\begin{equation} \label{eq:d_final}
    \bm d^{\ell}_i \approx  {\sum_{j \in \cN} \bbE\left[\bm Z_{i j}^{\ell}\right] } {= \sum_{j \in \cN} p_{ij}^{\ell}}, \; \forall i \in \cN.
\end{equation}

In the A-step, the aggregate topology from topologies of all layers in the multiplex network is obtained using the OR aggregation mechanism \cite{wu2022discrimination}. Specifically, the respective entries in the aggregate adjacency matrix should be equal to those calculated from the observed part of the adjacency matrix for each layer, 
which is expressed as

\begin{equation} \label{eq:agg_topo_final}
    {{\bm A}_{ij}} = \varphi(\bm Z_{ij}^{1}, \dots, \bm Z_{ij}^{|\cL|}) = 1 - \prod_{\ell \in \cL} (1 - \bm Z^{\ell}_{ij})
\end{equation}
\noindent
where $\varphi(\cdot)$ is the aggregation function. Apparently, $\bm A_{ij} =1$ when there exists at least one $\ell$ such that $\bm Z_{ij}^{\ell}=1$ and $\bm A_{ij} = 0$ only when $\bm Z_{ij}^{\ell}=0$ for all $\ell$. {The aggregate topology (where nodes are connected) is used to update the estimated complete topology of each layer as follows. According to the Bayes' Theorem, we have}

{
\begin{subequations} \label{eq:Z_given_A}
\begin{align}
    P(\bm Z_{ij}^{\ell} = 1 \mid \bm A_{ij}=1) &= \frac{P({\bm A_{ij} } 
     = 1 \mid \bm Z_{ij}^{\ell} = 1 ) \cdot P(\bm Z_{ij}^{\ell} = 1)}{P(\bm A_{ij}=1) } \\
     &= \frac{1 \cdot p_{ij}^{\ell} }{ 1 - \prod\limits_{h \in \cL} (1-p_{ij}^{h} )} \label{eq:p_given_A}
\end{align} 
\end{subequations}
}

\noindent
{In the implementation, this means that the latest $p_{ij}^{\ell}$ for all $i,j$, which is equivalent to $P(\bm Z_{ij}^{\ell} = 1 \mid \bm A_{ij}=1) $ at this point, is updated according to Eq. \eqref{eq:p_given_A}.}


The complete EMA algorithm is summarized in Algorithm \ref{alg:EMA}. Note that we can treat the aggregate topology as part of the parameters for generating the multiplex network, i.e., $\Theta=(\bm A, \bm d)$. From this perspective, the EMA framework is reduced to an EM framework and the A-step is regarded as a component of the M-step for estimating the optimal model parameters. This is because given the current estimate of the unobserved part of the multiplex network, the aggregate topology obtained by Eq. \eqref{eq:agg_topo_final} is the only solution and hence the optimal solution to the aggregate topology.

\begin{algorithm}
\caption{EMA for multiplex network reconstruction}\label{alg:EMA}
\begin{algorithmic}[1]
\Input{Error tolerance $\epsilon_T$, maximum number of iterations $Iter_{\max}$, number of unique nodes in the multiplex network $|\cN|$, and partially observed topology $\bm X$}.
\Output{Reconstructed complete topology $\bm Z$ for the multiplex network}.
\State Initialize the degree sequences and predicted adjacency matrices: $\bm d_i^{\ell} \sim U(1, |\cN|),\, \forall i \in \{1,\dots, |\cN|\}$ and $\bm Z_{ij}^{\ell} \sim U(0, 1),\, \forall i \in \{1,\dots, |\cN|\}$.
\For{$iter$ in 1 to $Iter_{\max}$}{}
    \State E-step: Update {$q (\bm Z)$} according to { Eq. \eqref{eq:q_Z} }. 
    \State A-step: 
    {Update $\bm Z$ using the aggregate topology according to Eq. \eqref{eq:Z_given_A} }.
    \State M-step: Update $\bm d$ according to Eq. \eqref{eq:d_final}. 
    \State Calculate {the error} $\epsilon$  
    \If{$\epsilon > \epsilon_T$ }{~Continue}  
    \Else{~Break}{}
    \EndIf
\EndFor
\end{algorithmic}
\end{algorithm}

\section{Datasets}

This section presents the details about the real-world multiplex network we use in validating the EMA framework. Covert multiplex networks include the drug trafficking multiplex network and the Sicilian Mafia multiplex network. To demonstrate that the EMA framework is general, we validate it against the other two models on overt multiplex networks, including the London transportation multiplex network and the C. elegans neural multiplex network. This set of networks provide an opportunity to test the EMA framework under varying fractions of observed components of multiplex networks and permit the comparison between social networks, biological, and physical networks.

\subsection{Drug Trafficking Multiplex Network} \label{sec:drug_net}

\begin{figure}[!ht]
    \centering
    \includegraphics[width=1.04\textwidth]{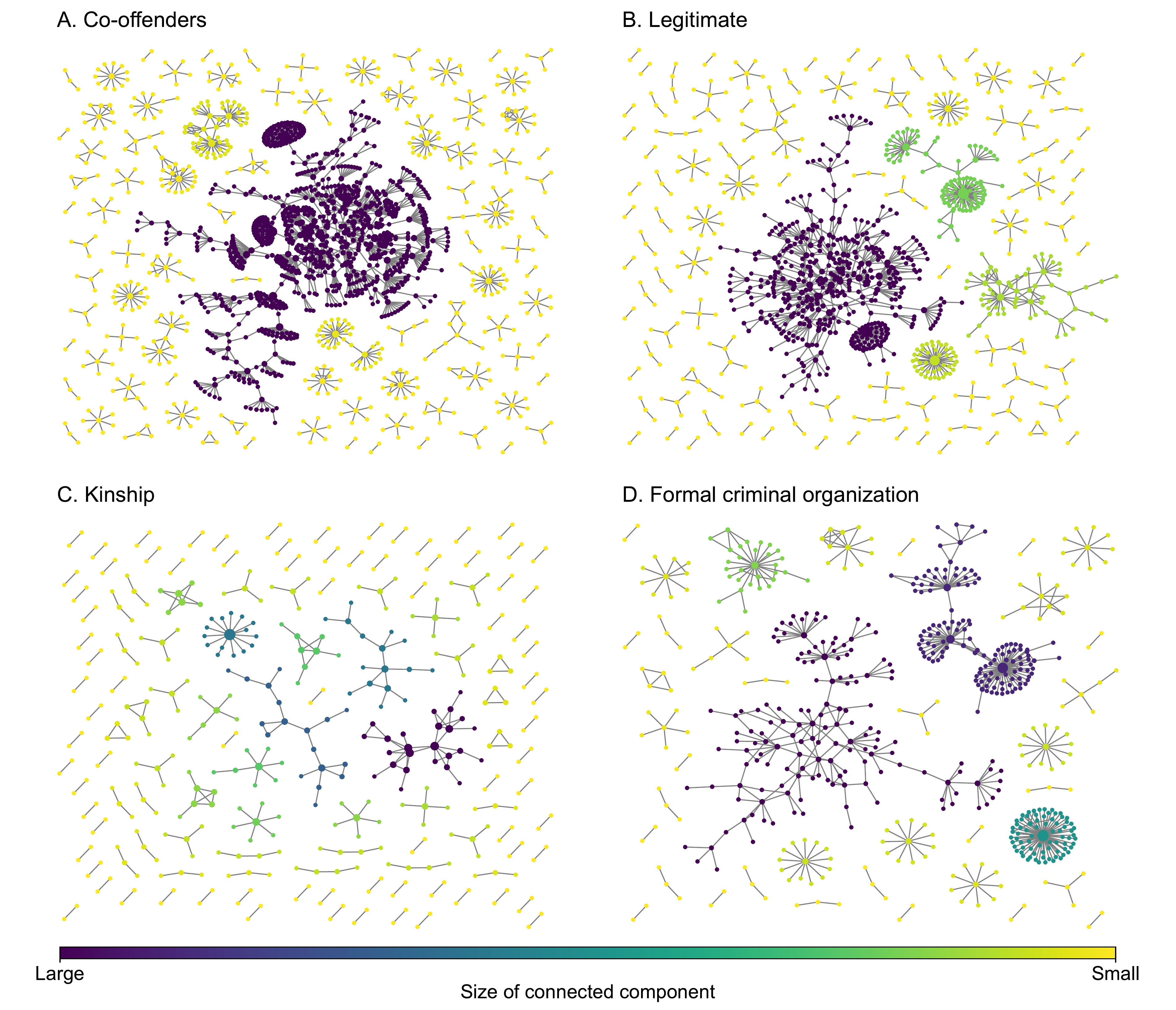} 
    \caption{Layers of the drug trafficking multiplex network. The size of a node is proportional to its degree.} 
    \label{fig:drug_layers}
\end{figure}

The dataset on drug trafficking networks in this study was obtained from the 2007 ‘E’ Division Provincial
Threat Assessment (PTA) report generated by Criminal Intelligence
Service Canada and the Royal Canadian Mounted Police (RCMP) \cite{malm2010comparing}. 
This report contains information about all individuals associated with the criminal operations of 129 different groups. This extensive data collection effort maps a community of drug trafficking operations, revealing how individuals interact within and between groups to sustain regional illicit drug supply and distribution. Individual-level information includes demographic characteristics,
description of three years of drug trafficking activity, their role in drug trafficking, and their current and historical relationships among each other, including co-offenders (individuals who commit crimes together), legitimate business partners (including friends), their kinship, and involvement in a formal criminal organization (including enemies) (Fig. \ref{fig:drug_layers}). This multiplex network integrated information obtained from official records management systems, surveillance, and non-criminal justice information sources. The layers for co-offenders and legitimate business partners are used to build a two-layer multiplex network (representing surveillance data capturing current activity, criminal or otherwise), the layers for co-offenders, legitimate business partners, and formal criminal organization are used to build a three-layer multiplex network (adding social embedding in a community of criminal actors), and the four-layer multiplex network adds familial relations (adding trusted relations that bolster network resiliency). It is obvious from the figure that both layers do not have giant hub nodes that are incident with over 50\% of the edges of a layer. 
Some key network statistics of each of the four layers are presented in Table \ref{tab:drug_layers}. We can see from the table that each layer has a low link density ($\rho$), low average degree ($\langle d \rangle $), and the average size of the connected component ($\langle |CC| \rangle$). Although the size of the greatest connected component is not small, particularly the one in the co-offender network, the coefficient of variance (CoV, i.e., ratio of standard deviation and mean) of the size of the connected component is small, indicating that members in drug trafficking community are generally isolated.


\begin{table}[!htb] 
  \centering
  \caption{Network statistics of each layer in the drug trafficking multiplex network. FCO is the shorthand for formal criminal organization.}
    \begin{tabular}{lccccccc}
    \toprule
    \multicolumn{1}{l}{Layer} & $|\cN|$   & $|\cE|$   & $\rho (\times 10^{-3})$   &  $\langle d \rangle $   & $\langle|CC|\rangle$ & $|GCC|$ & CoV of $|CC|$ \\
    \midrule
    Co-offender & 1645  & 1808  & 1.32  & 2.19  & 19.00 & 1024  & 5.71 \\
    Legitimate & 1022  & 1041  & 2.00     & 2.04  & 10.99 & 462   & 4.43 \\
    \multicolumn{1}{l}{FCO} & 560   & 597   & 3.81  & 2.13  & 14.36 & 150   & 2.24 \\
    Kinship & 399   & 308   & 3.88  & 1.54  & 3.22  & 25    & 0.99 \\
    \bottomrule
    \end{tabular}%
  \label{tab:drug_layers}%
\end{table}%

\subsection{Sicilian Mafia Multiplex Network} \label{sec:mafia_net}
\begin{figure}[!ht]
    \centering
    \includegraphics[width=0.95\textwidth]{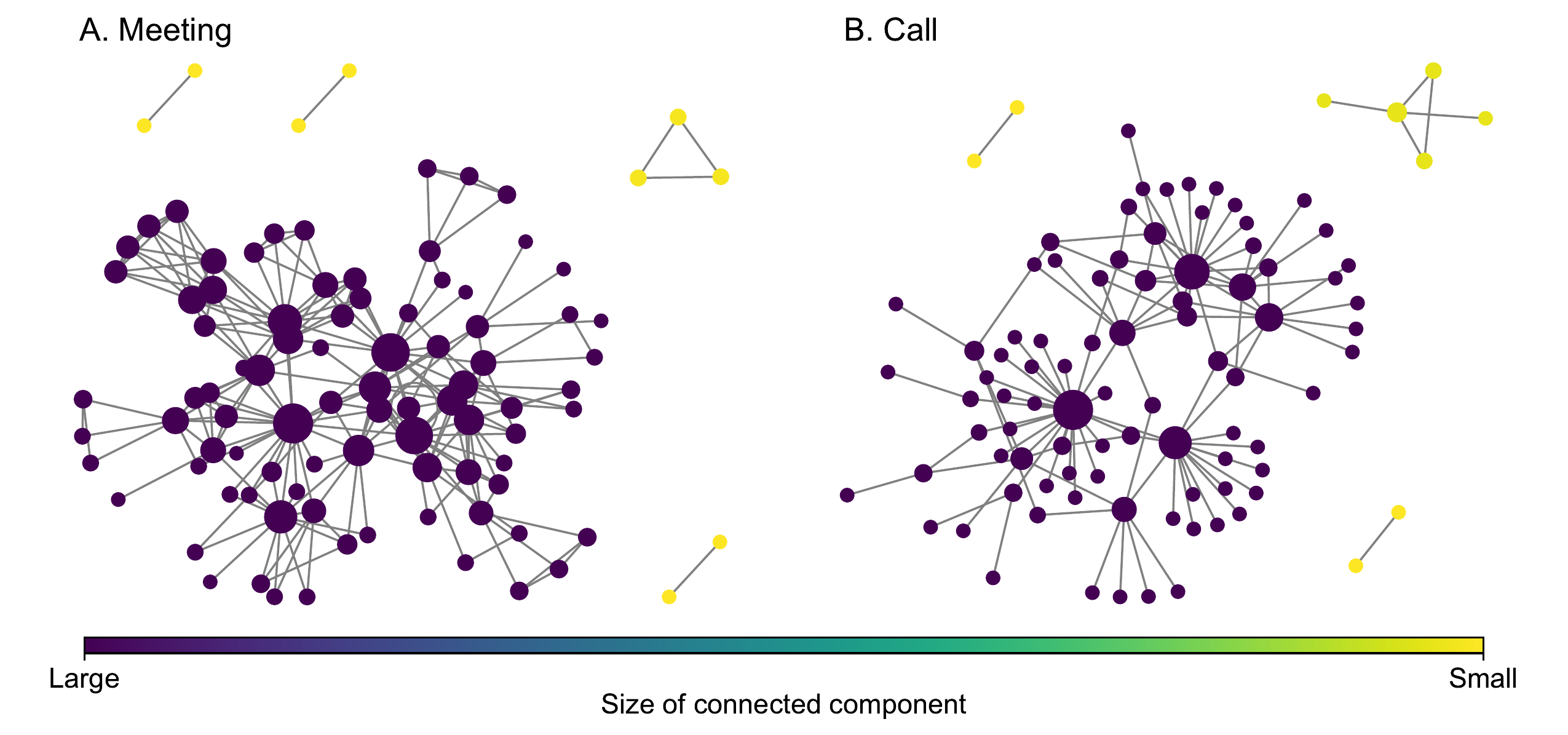} 
    \caption{Layers of the Sicilian Mafia multiplex network. The size of a node is proportional to its degree.}
    \label{fig:mafia_layers}
\end{figure}

The Sicilian Mafia multiplex network (Fig. \ref{fig:mafia_layers}) was derived from the pre-trial detention order issued by the local court in 2007 in response to a major investigation \cite{cavallaro2020disrupting}.The Italian civil law system is substantively different than federalist case law systems (such as the U.S. system), and as such, the court documents filed in support of the detention order provide sufficient information to map individual-level involvement in criminal enterprise activities \cite{calderoni2020robust}. Network information captures five years of observations (2003 to 2007) for two Mafia clans, known as the Mistretta family and the Batanesi clan. These familial clans are embedded within a community of illicit entrepreneurs collectively referred to as the Sicilian Mafia. This data set includes two types of layers: (i) meeting data among suspected (101 nodes and 256 edges) and (ii) phone calls among individuals (100 nodes and 124 edges). These layers represent different modes of communications commonly observed during surveillance activity in support of major investigations. Some key statistics are presented in Table \ref{tab:mafia_layers}. 
It can be observed from the figure that both layers do not have giant hub nodes as well.  

\begin{table}[!ht]
  \centering
  \caption{Network statistics of each layer in the Sicilian Mafia multiplex network}
    \begin{tabular}{lccccccc}
    \toprule
    \multicolumn{1}{l}{Layer} & $|\cN|$   & $|\cE|$   & $\rho (\times 10^{-3})$   &  $\langle d \rangle $   & $\langle|CC|\rangle$ & $|GCC|$ & CoV of $|CC|$ \\
    \midrule
    Meeting & 101   & 256   & 55.77 & 5.07  & 19.02    & 86    & 1.76 \\
    Call  & 100   & 124   & 27.45 & 2.48  & 23.50  & 85    & 1.51 \\
    \bottomrule
    \end{tabular}%
  \label{tab:mafia_layers}%
\end{table}%

\subsection{London Transportation Multiplex Network}

\begin{figure}[!ht]
    \centering
    \includegraphics[width=1.04\textwidth]{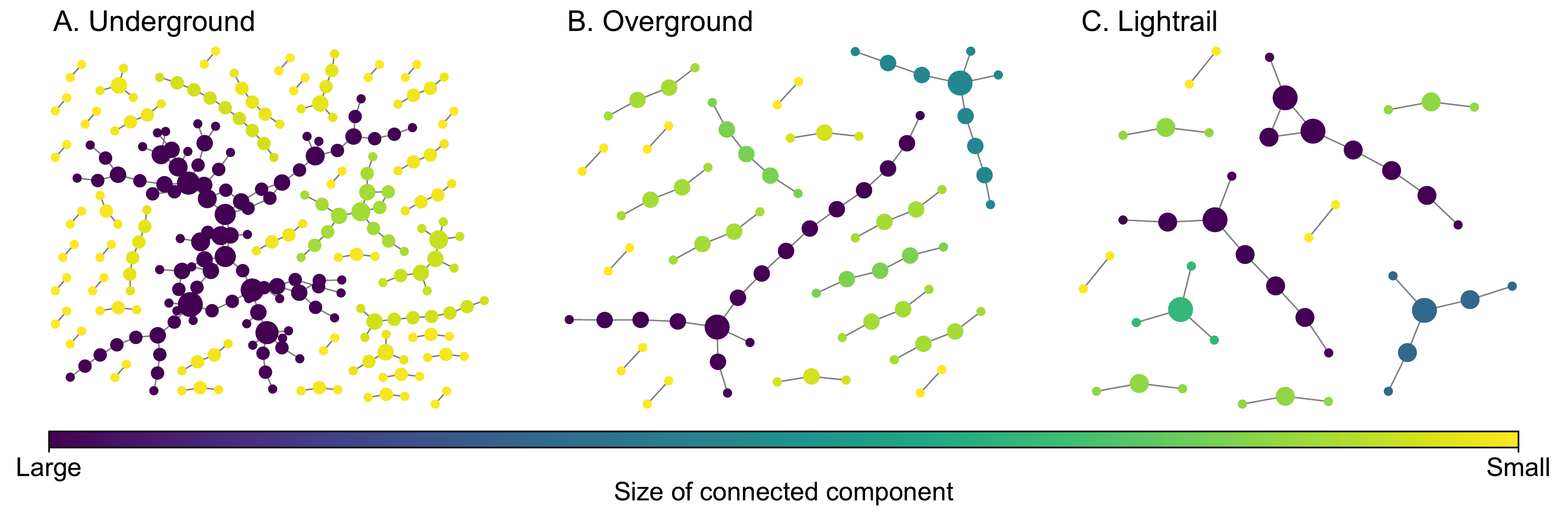} 
    \caption{Layers of the London transportation multiplex network. The size of a node is proportional to its degree.}
    \label{fig:transport_layers}
\end{figure}

The London transportation multiplex network \cite{de2014navigability} (Fig. \ref{fig:transport_layers}) consists of underground (260 nodes and 225 edges), overground (81 nodes and 62 edges), and lightrail (44 nodes and 34 edges). The overground and light rail layers are selected to construct a two-layer multiplex network. The third layer adds the underground edges. While the modalities (layers) of transit systems are designed to intersect forming a fully connected giant component, single layers will not always form a connected component. Moreover, this network is intentionally incomplete in that buses that may link different modalities (e.g., underground to light rail) were not included. Each layer of this transportation multiplex network has many connected components and the size of the greatest connected component is not very large compared to the number of nodes in each layer.

\begin{table}[!ht]
  \centering
  \caption{Network statistics of each layer in the London transportation multiplex network}
    \begin{tabular}{lccccccc}
    \toprule
    \multicolumn{1}{l}{Layer} & $|\cN|$   & $|\cE|$   & $\rho (\times 10^{-3})$   &  $\langle d \rangle $   & $\langle|CC|\rangle$ & $|GCC|$ & CoV of $|CC|$ \\
    \midrule
    Underground & 260   & 225   & 6.68  & 1.73  & 5.65  & 98    & 2.48 \\
    Overground & 81    & 62    & 19.14 & 1.53  & 4.26  & 17    & 0.83 \\
    Lightrail & 44    & 34    & 35.94 & 1.55  & 4.00     & 8     & 0.54 \\
    \bottomrule
    \end{tabular}%
  \label{tab:transport_layers}%
\end{table}%

\subsection{C. Elegans Multiplex Network}
\begin{figure}[!ht]
    \centering
    \includegraphics[width=1.04\textwidth]{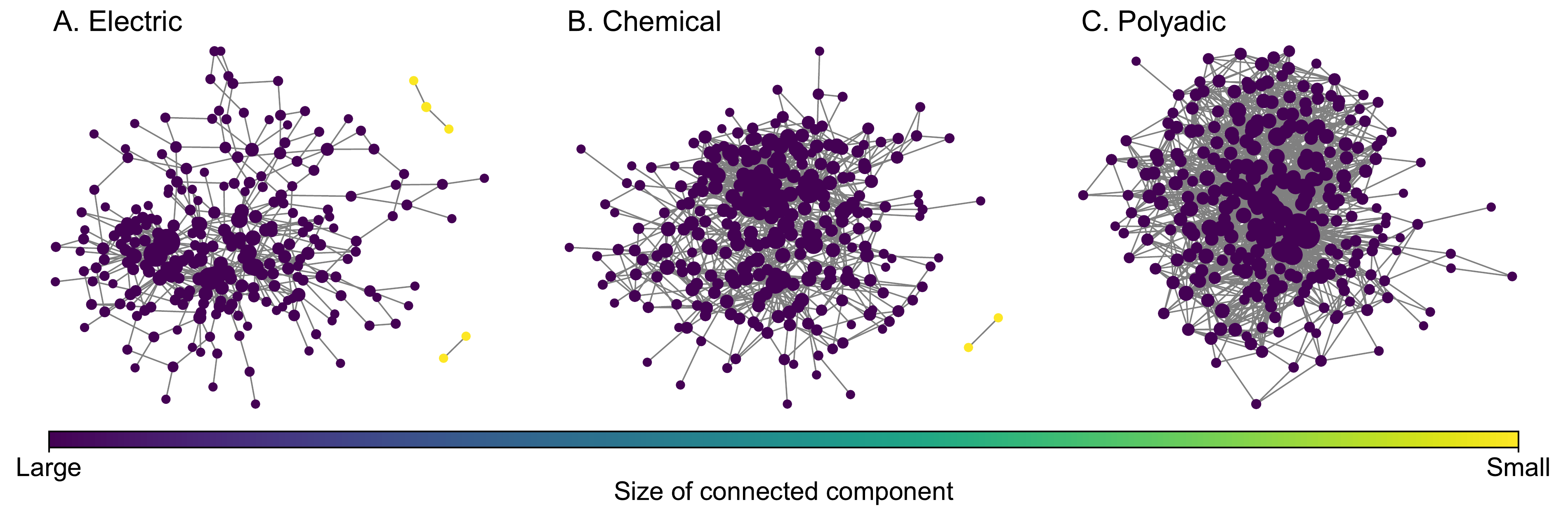}
    \caption{Layers of the C. elegans multiplex network. The size of a node is proportional to its degree.}
    \label{fig:elegans_layers}
\end{figure}

The C. (Caenorhabditis) elegans neural multiplex network \cite{chen2006wiring} (Fig. \ref{fig:elegans_layers}) consists of three layers for different types of synaptic junctions, including electric (242 nodes and 450 edges), chemical (260 nodes and 869 edges), and polyadic (277 nodes and 1666 edges). Layers for chemical and polyadic junctions are selected to construct a two-layer multiplex network. Compared to the covert multiplex networks introduced earlier, all layers of the C. elegans multiplex network are much denser because each layer has a higher density. In each layer, the size of the greatest connected component is large. In particular, all the nodes in the polyadic layer are connected because the size of the greatest connected component is equal to the number of nodes.

\begin{table}[!ht]
  \centering
  \caption{Network statistics of each layer in the C. elegans multiplex network}
    \begin{tabular}{lccccccc}
    \toprule
    \multicolumn{1}{l}{Layer} & $|\cN|$   & $|\cE|$   & $\rho (\times 10^{-3})$   &  $\langle d \rangle $   & $\langle|CC|\rangle$ & $|GCC|$ & CoV of $|CC|$ \\
    \midrule
    Electrical & 242   & 450   & 15.43 & 3.72  & 80.67 & 237   & 1.37 \\
    Chemical & 260   & 869   & 25.81 & 6.68  & 130.00   & 258   & 0.98 \\
    Polyadic & 277   & 1666   & 43.58 & 12.03  & 277.00   & 277   & 0.00 \\
    \bottomrule
    \end{tabular}%
  \label{tab:elegans_layers}%
\end{table}%

\section{Experiments}

\subsection{Experiment Setup}

For each multiplex network, we use 50 random simulations at each fraction of missing nodes to obtain the average predictive performance (we find that 50 repetitions can already achieve a stable mean value). When a node is not observed, the links connected to this node are also considered unobserved. For both EMA and EM, the error tolerance $\epsilon_T$ is set to $1 \times 10^{-5}$, i.e., the solution is considered converged if the difference in the mean absolute error (MAE) of the predicted adjacency matrices (unobserved entries only) at two consecutive iterations for all unobserved links is lower than the tolerance.

Because covert networks are sparse, the associated datasets are typically imbalanced toward negatives (no links between nodes), we employ the Matthews Correlation Coefficient (MCC) and G-mean 
to measure the prediction (classification) accuracy \cite{branco2016survey,boughorbel2017optimal}. MCC can be computed as 



\begin{equation}
    \textrm{MCC}=\frac{\textrm{TP} \times \textrm{TN}-\textrm{FP} \times \textrm{FN}}{\sqrt{(\textrm{TP}+\textrm{FP}) \times(\textrm{TP}+\textrm{FN}) \times(\textrm{TN}+\textrm{FP}) \times(\textrm{TN}+\textrm{FN})}}
\end{equation}
\noindent where TP represents true positives, TN represents true negatives, FP represents false positives, and FN represents false negatives. Note that the value of MCC can be negative. G-mean is given by

\begin{equation}
    \textrm{G-mean} =\sqrt{\frac{\textrm{TP}}{\textrm{TP}+\textrm{FN}} \times \frac{\textrm{TN}}{\textrm{TN}+\textrm{FP}}} = \sqrt{\textrm{Recall} \times \textrm{Specificity}}
\end{equation}

We also include the random model (RM), i.e., a binary uniformly random classifier, as a baseline model. Among all possible links except for the links among observed nodes, the RM selects a number (i.e., the number of links left) of links in a uniformly random manner as the predicted links.

\subsection{Results}

\begin{figure}[!ht]
    \centering
    \includegraphics[width=0.65\textwidth]{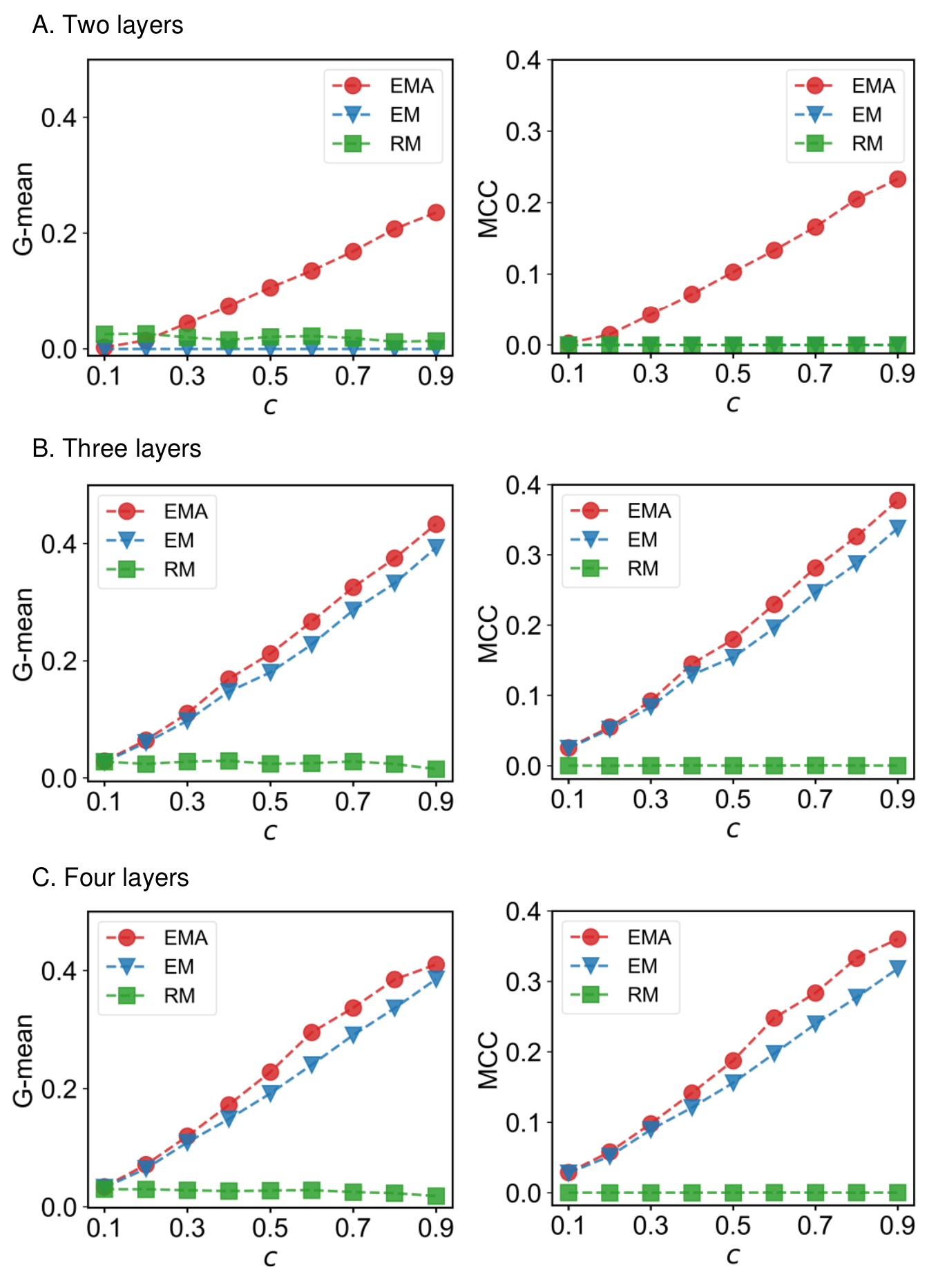}
    \vspace{-4pt}
    \caption{G-mean and MCC for drug trafficking networks}
    \label{fig:drug_accuracy}
\end{figure}

\begin{figure}[!ht]
    \centering
    \vspace{4pt}
    \includegraphics[width=0.65\textwidth]{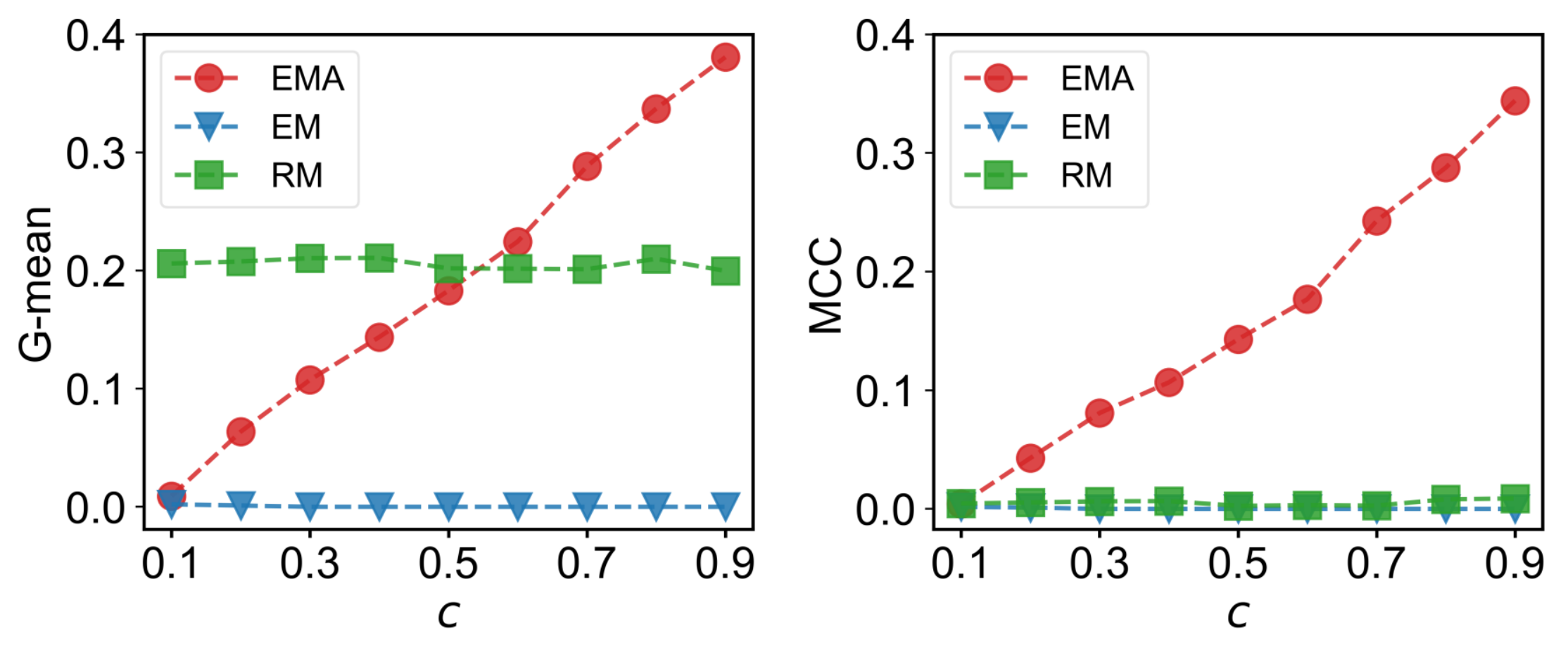}
    \vspace{-4pt}
    \caption{G-mean and MCC for Sicilian Mafia networks}
    \label{fig:mafia_accuracy}
\end{figure}

\begin{figure}[!ht]
    \centering
    \includegraphics[width=0.65\textwidth]{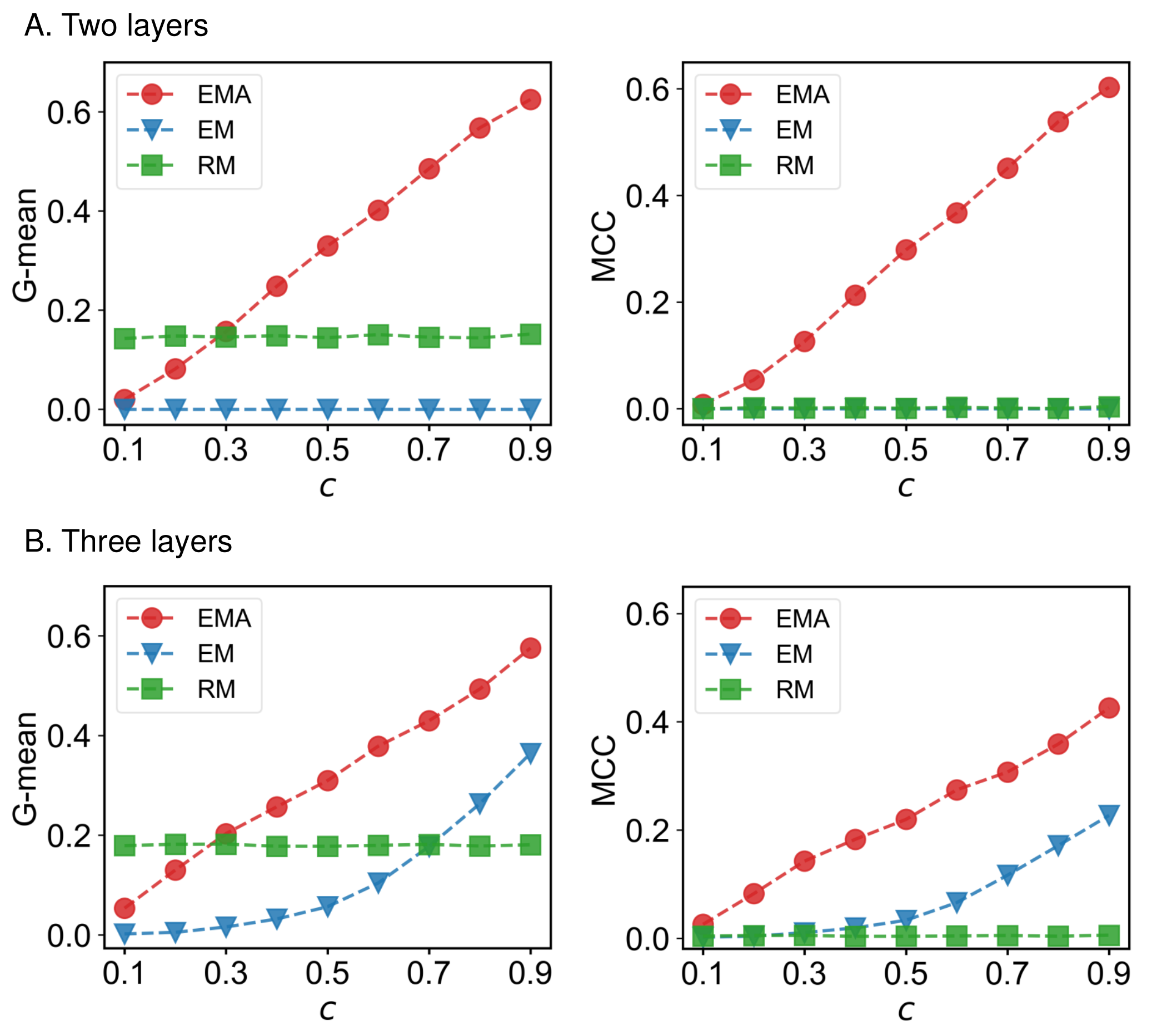}
    \vspace{-4pt}
    \caption{G-mean and MCC for C. elegans neural networks}
    \label{fig:elegan_accuracy}
\end{figure}

\begin{figure}[!h]
    \centering
    \vspace{4pt}
    \includegraphics[width=0.66\textwidth]{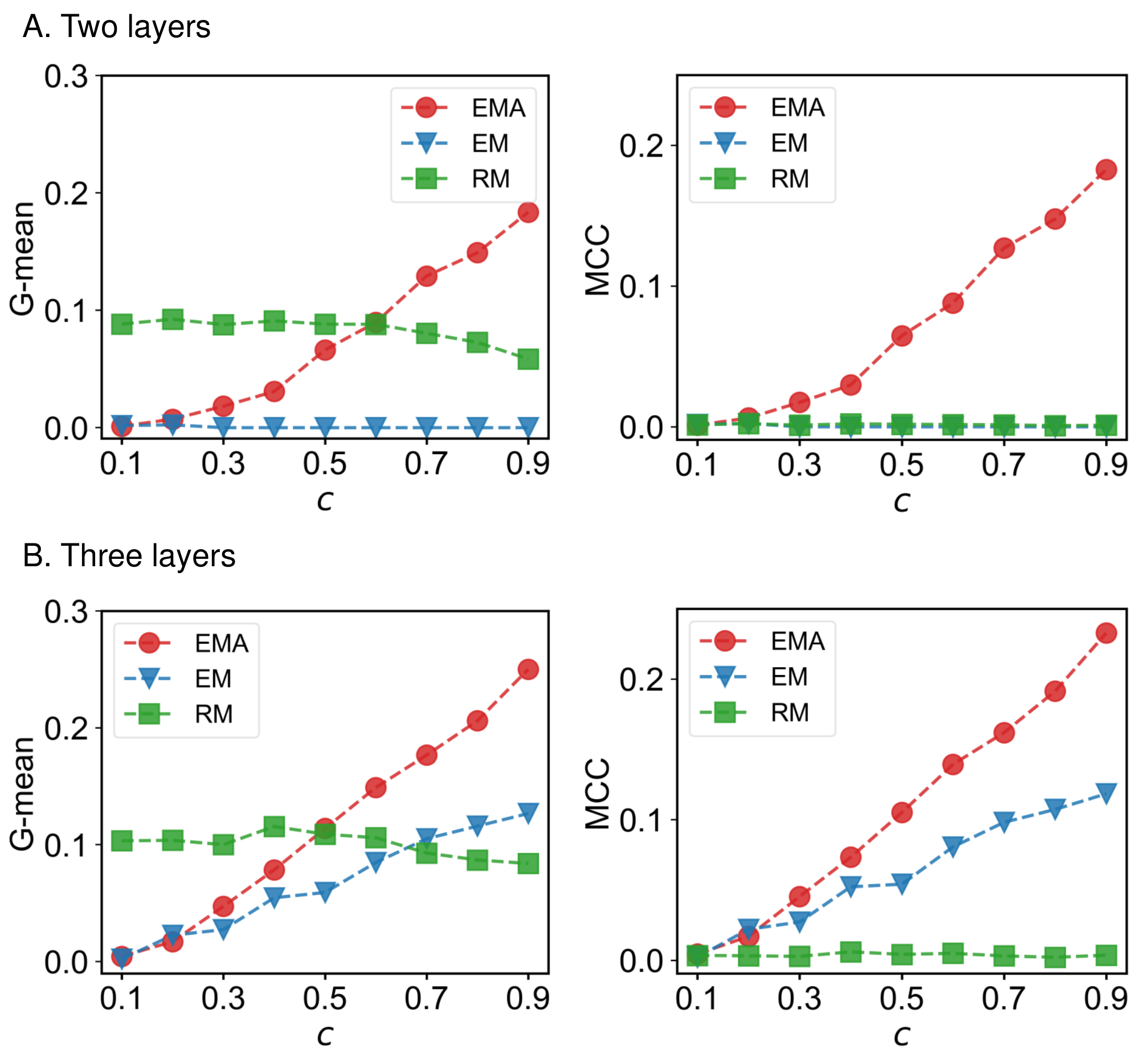}
    \vspace{-4pt}
    \caption{G-mean and MCC for London transportation networks}
    \label{fig:transport_accuracy}
\end{figure}

We first check the performance of EMA under different fractions of missing components. The predictive performances measured by G-mean and MCC for different multiplex networks under 10\% to 90\% of missing components are presented in Fig. \ref{fig:drug_accuracy} to Fig. \ref{fig:transport_accuracy} where $c$ represents the fraction of {observed} components. We can see that EMA achieves the best MCC regardless of the multiplex networks and the fraction of missing components. In terms of G-mean, EMA outperforms EM in all cases and random model when the observed fraction of components is large. Note that the trends of G-mean and MCC over different fractions of observed components obtained by EMA and EM are close to each other because the values of FP $\times$ FN are much smaller than TP $\times$ TN.

Next, we examine the impact of the number of layers on the EMA and EM approaches. As the number of layers increases from two to three in drug trafficking (Fig. \ref{fig:drug_accuracy}), C. elegans (Fig. \ref{fig:elegan_accuracy}), and London transport networks (Fig. \ref{fig:transport_accuracy}), the performance improvement of EMA when compared to EM in both G-mean and MCC decreases, indicating that the benefits of adding the aggregation step drops. This is because when there are more layers, there are more choices of values for individual entries in the layers, making a positive entry in the aggregate topology less informative about the positive value of each layer.

\begin{figure}[!ht]
    \centering
    \includegraphics[width=1.01\textwidth]{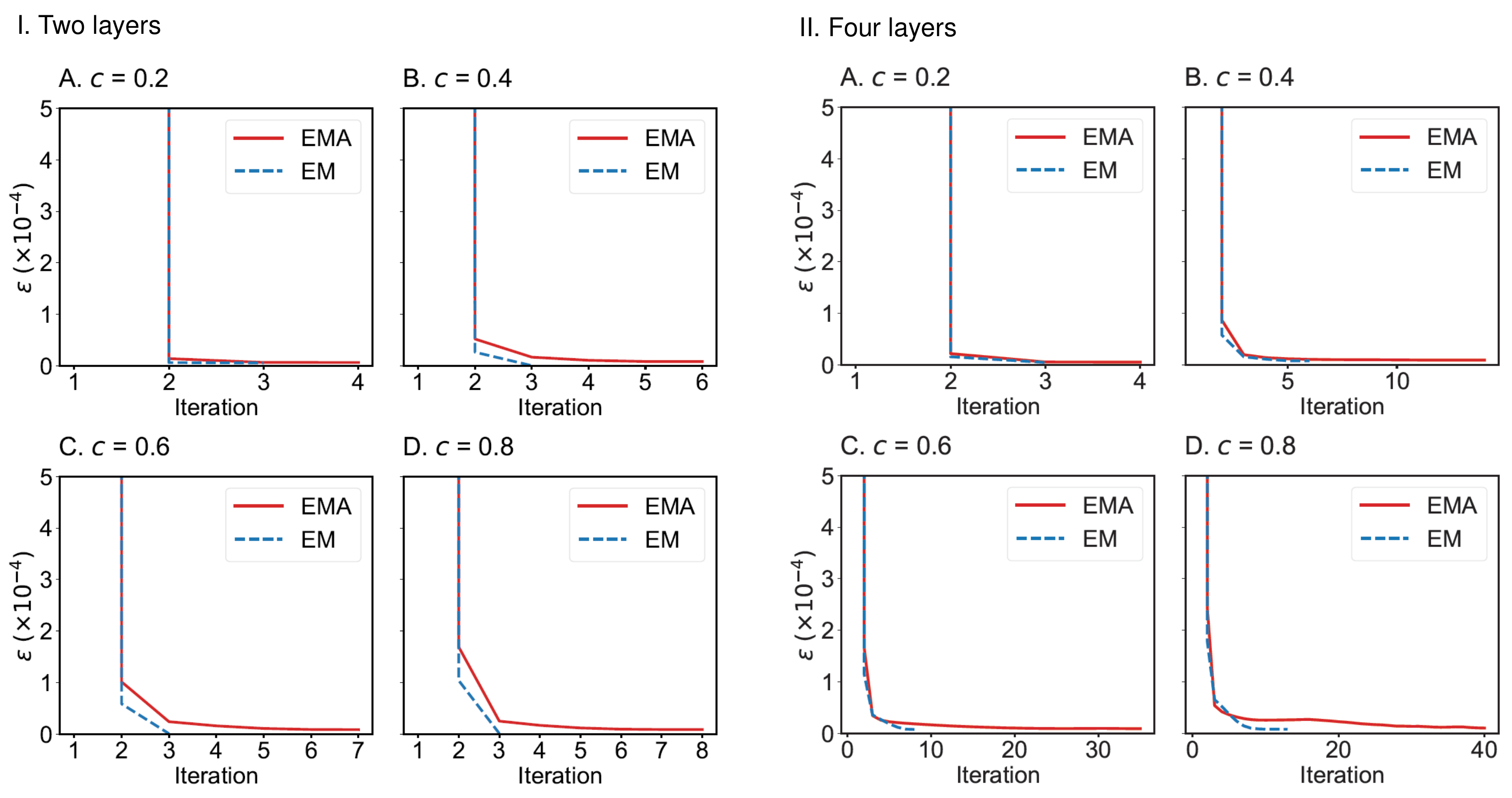}
    \caption{Convergence of EMA and EM on the two-layer (I) and four-layer (II) drug trafficking multiplex networks under different fractions of observed components: A. 0.2; B. 0.4; C.0.6; D. 0.8.}
    \label{fig:drug_converge}
\end{figure}

\begin{figure}[!ht]
    \centering
    \includegraphics[width=1.01\textwidth]{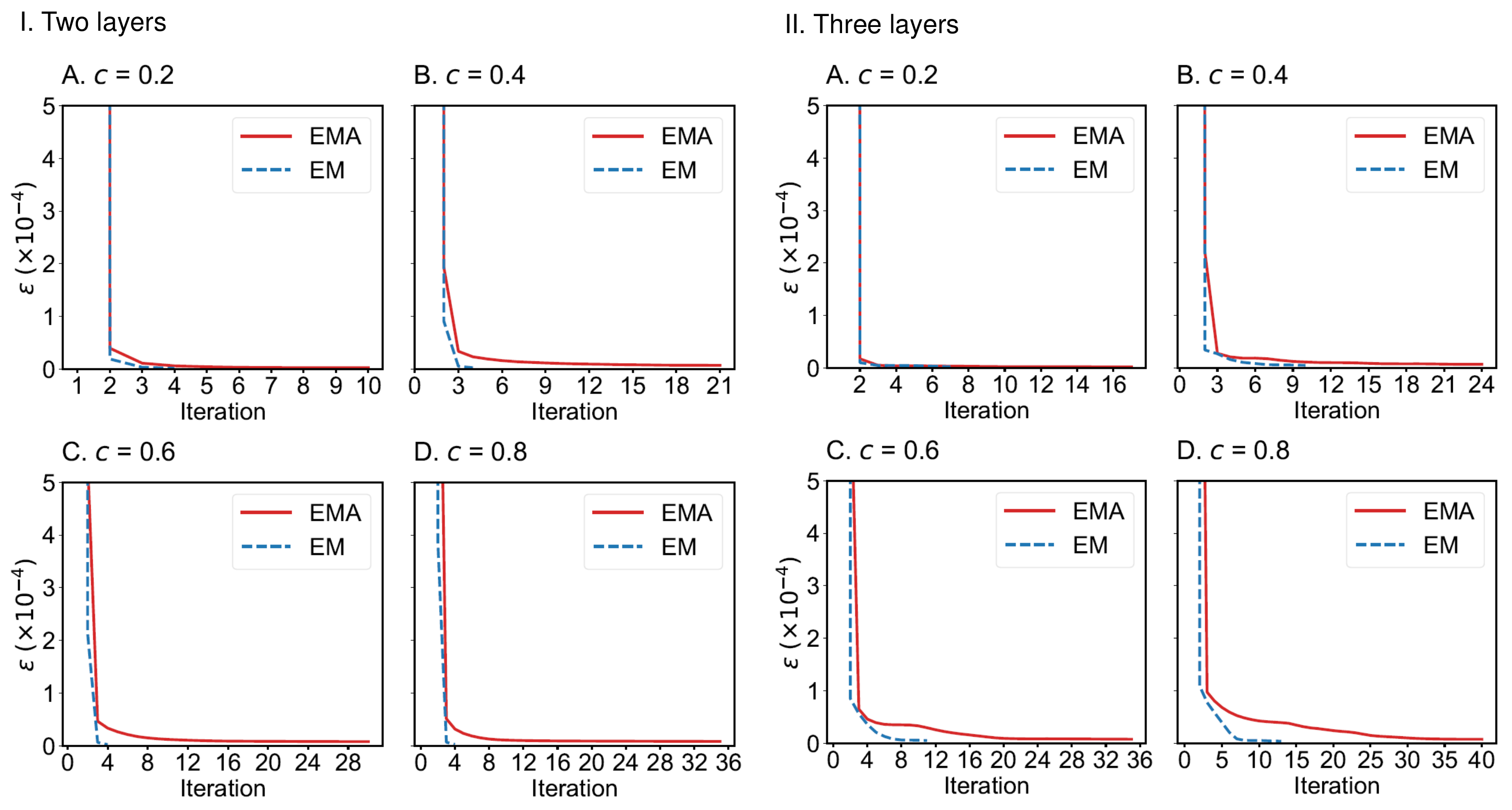}
    \caption{Convergence of EMA and EM on the two-layer (I) and three-layer (II) London transportation multiplex networks under different fractions of observed components: A. 0.2; B. 0.4; C.0.6; D. 0.8.}
    \label{fig:transport_converge}
\end{figure}

We also compare the convergence of EMA and EM algorithms on the drug trafficking multiplex networks (Fig. \ref{fig:drug_converge}) and London transportation multiplex networks (Fig. \ref{fig:transport_converge}). Regardless of the size of multiplex networks and the fraction of components observed, EM converges more quickly than EMA. In the worst-case tested, EMA can reach convergence within 40 iterations given a convergence error of $1 \times 10^{-5}$. However, if the convergence error can be set to a larger value, such as $1 \times 10^{-4}$, then EMA can achieve convergence much faster (within 10 iterations across the networks considered). Therefore, both algorithms can be applied to moderately large-scale multiplex networks. 

\section{Discussion}

From a criminal justice perspective, mapping networks using available data on illicit activities is typically not sufficient to obtain the complete structure of the covert networks under study. 
As a result, networks of criminal operations are always incomplete, thereby compromising the potential practical contributions network science can make toward developing effective interdiction strategies.  
Two major implications for Disrupting Illicit Networks arise from this investigation for reconstructing a more complete structure of multiplex networks. 

First, with regard to the broad context of disrupting covert networks, the experiments support arguments that criminal enterprise is intimately embedded within a complex social system wherein a pair of individuals can be connected by multiple relations or a criminal actor could tap into different social networks to resolve a pressing issue \cite{morselli2006legitimate, malm2013using, diviak2020poisonous}. For example, a drug smuggler uses legitimate business assets (helicopter and employed pilot) to transport illicit products (multiple relations) and in a pinch, the smuggler could launder proceeds through a relative's real-estate business (temporary use of a different type of relation for criminal ends). The extent to which associates and kin are aware of these illegalities is not necessarily relevant; what is critical is that this interwoven social fabric must be fully understood before effective interdiction strategies can be developed and implemented. Results on the performance of the EMA framework for the drug trafficking network and the Sicilian mafia network show that networks generated from the surveillance of current criminal activity are insufficient. Interpolating latent connections by adding at least one more layer that captures historic criminogenic relations (i.e., formal organized crime membership), business or professional associations, or purely social and familial connections, significantly improves network completeness. This finding coincides with applied network criminology \cite{malm2010comparing, smith2016trust} and crime opportunity theory \cite{bichler2020understanding, kleemans2008criminal}.

Second, while inference can improve network completeness, such modeling frameworks do not constitute a silver bullet for all methodological issues. Continued efforts are needed to improve data collection protocols to enhance the capacity to generate more complete information about covert operations and the complex social systems within which they are embedded. To this aim, our experiments show that identifying additional layers composing a multiplex network makes the EMA algorithm more accurate. 
Supporting prior arguments to establish methodological conventions regarding the use of criminal justice system data in network applications \cite{campana2012listening, diviak2019key}, this finding confirms that the key to uncovering more of the covert multiplex network is to tap different layers of information sources \cite{smith2016trust, bichler2020understanding}. However, additional tests are needed to understand the nature of information benefits. For example, 
How does the dependency between layers, particularly in social networks, impact information benefits? When does identifying additional layers of covert networks benefit more than identifying more components within known layers? Finally, given that the covert networks were mapped from observations made over many years, does the model offer more information benefits when used with shorter observations, such as six months?


\section{Conclusions}
In this study, we develop the EMA framework for reconstructing multiplex networks and validate this approach against the EM approach and the random model on several multiplex networks given different fractions of the observed part of the complete networks. We find that EMA achieves the best predictive performance in terms of G-mean and MCC over different levels of available observed parts of the true networks. The findings are consistent across experiments conducted on different types of multiplex networks (social and non-social). This indicates that integrating the estimation of the aggregate topology, which encodes the interlayer dependency of multiplex networks, can improve the reconstruction accuracy. By inferring a more complete structure of covert multiplex networks, the EMA framework can support interdiction efforts aimed at disrupting criminal operations, especially when the observations are only available for two or three types of interactions. As the fraction of observed actors and the type of interactions among actors increases, the added benefits of the EMA framework when compared to the EM framework decrease. In cases where quick decisions are required for interdicting multiplex with many layers, the EM framework is recommended due to a lower computational load.

Future work will consider the use of graph neural networks to reconstruct sparse multiplex networks. Although GNN has been widely applied to predict missing links \cite{trouillon2016complex,zhang2018link,wang2019robust}, research efforts that apply GNN to reconstruct multiplex networks are limited. 
As such, it is valuable to compare the reconstruction accuracy of EMA and GNN on multiple datasets and identify where each approach performs the best. Also, since the accuracy of the EMA framework reduces for multiplex networks with many layers, future research can explore how to improve the predictive accuracy using network generative models with a prescribed level of interlayer dependency \cite{nicosia2013growing,yu2020modeling,wang2021generating,fugenschuh2021structural}, and use the dynamics of the networks \cite{jiang2020inferring}.


\section*{Data and Code Availability Statement}

Data sharing is not applicable to this paper since no new data were created. The code and datasets used for the numerical experiments are available at \texttt{\url{https://github.com/jinzhuyu/multiplex_recon}}.



\bibliographystyle{elsarticle-num-names}
\bibliography{Main_references.bib}
\end{document}